\documentclass[twocolumn]{aanomarks}  
\usepackage{graphicx}
\usepackage{natbib}
\usepackage{color}           % For color text: \color command
\definecolor{mgc}{RGB}{0,0,0}

\definecolor{mrk}{RGB}{0,0,160}

\definecolor{dltd}{RGB}{200,100,100}

\newcommand{\ang}{\mathrm{\AA}}

\begin{document}

\title{Using the Stokes V widths of Fe I lines for diagnostics of the intrinsic solar photospheric magnetic field}

   \author{M. Gordovskyy \inst{1}\fnmsep\thanks{\email{mykola.gordovskyy =AT= manchester.ac.uk}}, 
	  S. Shelyag\inst{2}, 
	P.K. Browning\inst{1}     
          \and
	V.G. Lozitsky\inst{3} 
	}

   \institute{Jodrell Bank Centre for Astrophysics, University of Manchester, Manchester M13\,9PL, UK
          \and
Faculty of Science, Engineering and Built Environment, Deakin University, Melburn VIC 3125, Australia
         \and
             Astronomical Observatory of the Taras Shevchenko National University of Kyiv, Observatorna 3, Kyiv 04053, Ukraine
             }

\authorrunning{Gordovskyy et al.}
\titlerunning{Using Stokes V widths for the photospheric magnetic field diagnostics}

% \abstract{}{}{}{}{} 
% 5 {} token are mandatory
 
  \abstract
   {}
   {The goal of this study is to explore a novel method for the solar photospheric magnetic field diagnostics using Stokes V widths of different magnetosensitive Fe~I spectral lines.}
   {We calculate Stokes I and V profiles of several Fe I lines based on a one-dimensional photospheric model VAL C using the NICOLE radiative transfer code. These profiles are used to produce calibration curves linking the intrinsic magnetic field values with the widths of blue peaks of Stokes V profiles. The obtained calibration curves are then tested using the Stokes profiles calculated for more realistic photospheric models based on MHD models of magneto-convection.}
   {It is shown that the developed Stokes V widths (SVW) method can be used with various optical and near-infrared lines. Out of six lines considered in this study, Fe~I~6301 line appears to be the most effective: it is sensitive to fields over $\sim$200~G and does not show any saturation up to $\sim$2~kG. Other lines considered can also be used for the photospheric field diagnostics with this method, however, only in narrower field value ranges, typically from about 100~G to 700--1000~G.}
   {The developed method can be a useful alternative to the classical magnetic line ratio method, particularly when the choice of lines is limited.}


   \maketitle
%
%________________________________________________________________

\section{Introduction}\label{s-intro}

Observations show that the solar photospheric magnetic field is inhomogeneous at spatial scales as short as 10~km, which are not resolved in most optical solar observations. Various observational and computational studies show that the photospheric magnetic flux is formed by small-scale magnetic fluxtubes with field strength of a few kG surrounded embedded into a much weaker ($\sim$10--100~G) ambient field \cite[see][for review]{frst72,sola93,dewe09,sten11,beor19}. 

\begin{figure*}[ht!]    
\centerline{\includegraphics[width=0.65\textwidth,clip=]{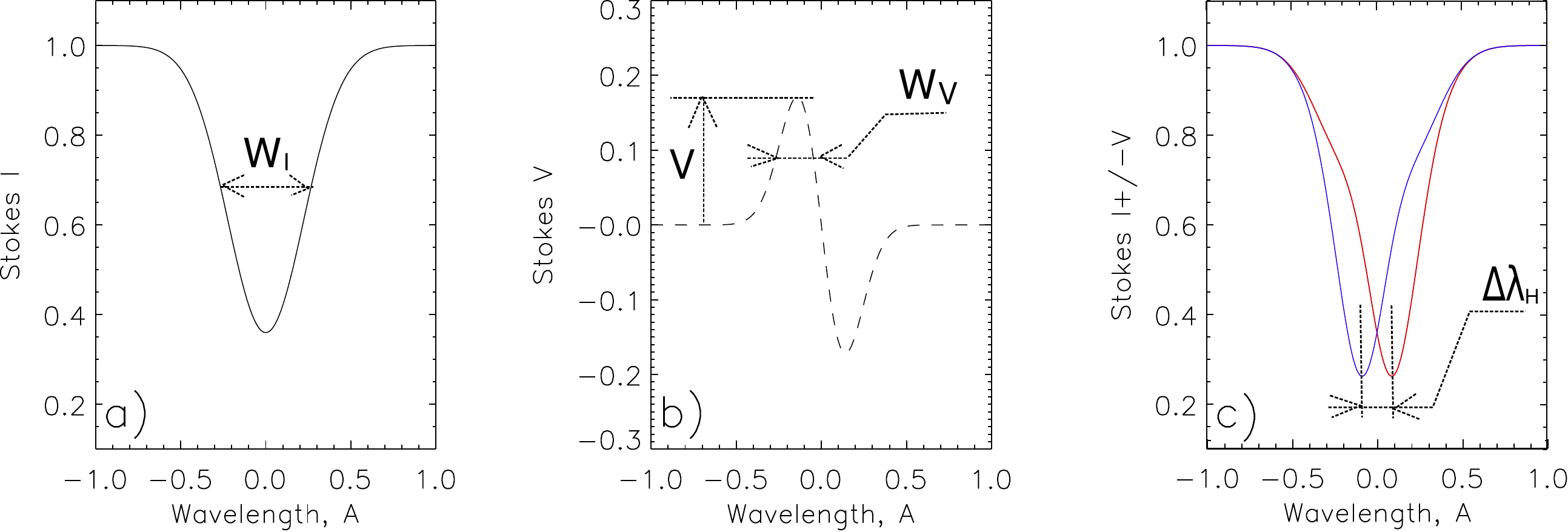}}
\caption{Sketch of line profiles, showing Stokes I profile (panel a), Stokes V (panel b), and Stokes I$\pm$V profiles (panel c), indicating the parameters used in this study.}
\label{f-sketch}aanomarks
\end{figure*}

\begin{figure*}[ht!]    
\centerline{\includegraphics[width=0.75\textwidth,clip=]{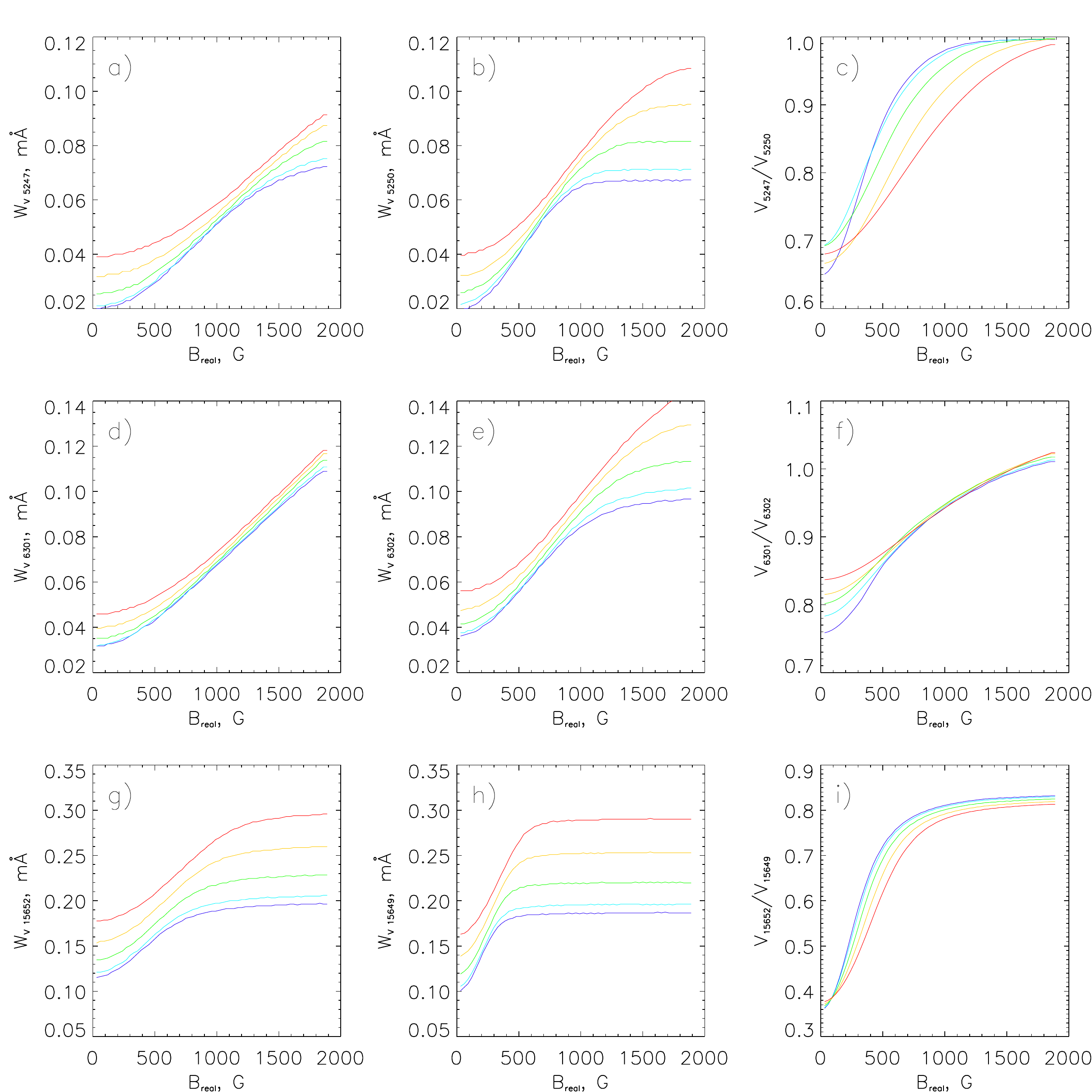}}
\caption{Calibration curves showing Stokes V width and magnetic line ratio values as functions of the intrinsic field $B$ for different turbulent velocities, which determine $W_I$. The turbulent velocities vary from 0 (blue lines) to 20~km/s (red lines). Panels (a-c), (d-f) and (g-i) correspond to 5247/5250, 6301/6302 and 15652/15649 pairs, respectively. The left column (panels a, d and g) show Stokes V widths for lines with lower Lande factors $g$, the middle column (panels b, e and h) show Stokes V widths for lines with higher Lande factors $g$, and the right column (panels c, f and i) shows MLR values.}
\end{figure*}

The photospheric magnetic field is normally measured using the Zeeman effect, through Zeeman splitting of the sigma-components of spectral lines or through the amplitudes of Stokes V profiles. In the case of spatially-uniform magnetic fields, the measured values represent the longitudinal component of the field vector. However, when the magnetic field is not uniform on the scales not resolved by the instrument, the measured values also depend on the magnetic filling factor, the magnetic field gradient along the line-of-sight (LOS) and other factors which affect the profiles of spectral lines. In the first approximation, known as the two-component model with zero ambient field, the observed field values $B_\mathrm{obs}$ are determined by the actual (or intrinsic) magnetic field, corresponding to the small-scale fluxtubes $B_\mathrm{real}$, and the filling factor $\alpha$:

\[ 
B_\mathrm{obs} = \alpha B_\mathrm{real}.
\]

Knowing the photospheric magnetic field is essential in many areas of solar physics, from observational helioseismology to the coronal magnetic field reconstruction. Some applications are not sensitive to the small-scale structure of the magnetic field. For instance, force-free coronal magnetic field reconstruction is normally sensitive only to the large-scale distribution of photospheric magnetic field \cite[e.g.][]{wisa12}. At the same time, there are characteristics which are very sensitive to the magnetic field filling factor. For instance, the Poynting flux and the magnetic energy density at the photosphere are often used for observational studies of the coronal heating and coronal magnetic field stability  \citep{wels14,jang16}. These values are proportional to $\alpha B_\mathrm{real}^2$ per pixel or, using the observed field values, $B_\mathrm{obs}^2 /\alpha$. Another example is evaluation of the electric current and Ohmic dissipation in the photosphere and chromosphere \cite[e.g.][]{muse15}. The current density is proportional to the spatial gradient of the magnetic field, which, in turn is proportional to $B_\mathrm{obs}/\alpha$. Hence, the energy deposition rate due to Ohmic heating should be also proportional to $B_\mathrm{obs}^2 /\alpha$. (However, in this case energy dissipation would also depend on actual sizes of the small-scale magnetic elements, electric resistivity and other parameters.) Thus, ignoring the unresolved field structure (i.e. assuming $\alpha=1$) would lead to underestimation of the photospheric magnetic energy and of the Ohmic heating by factor of $1/\alpha$. Therefore, evaluating the intrinsic magnetic field (or the filling factor) is one of the essential problems in interpreting magnetographic observations. 

\begin{table*}[ht!]
\caption{Atomic parameters of the six considered lines: wavelength, Lande factor, excitation potential of the lower level, transition probability, lower and upper term configurations.}
\begin{center}
\begin{tabular}{ l l l l l l }
\hline
    $\lambda$, $\ang$ & $g$ & $\chi$, eV & log($gf$) & Lower term & Upper term\\ 
\hline
    5247.1 & 2.00 & 0.087 & -4.97 & 5D2 & 7D3 \\
    5250.2 & 3.00 & 0.121 & -4.96 & 5D0 & 7D1 \\
    6301.5 & 1.67 & 3.654 & -0.72 & 5P2 & 5D2 \\
    6302.5 & 2.50 & 3.686 & -1.24 & 5P1 & 5D0 \\
    15648.5 & 3.00 & 5.426 & -0.043 & 7D5 & 7D4 \\
    15652.9 & 1.53 & 6.246 & -0.675 & 7D1 & 7D1 \\
\hline
\end{tabular}
\end{center}
\end{table*}

The most comprehensive method of the magnetic filling factor (or intrinsic magnetic field) diagnostics is the Stokes inversion \cite[see e.g.][for review]{sona01,toru16}. However, this method is computationally expensive and is usually applied to relatively small photospheric areas, consisting of up to $10^3$--$10^4$ pixels \cite[e.g.][]{kabo19,kuck19}. Hence, it cannot be realistically used to analyse big photospheric area, such as a large active region ($10^4$--$10^6$ pixels), or the whole Sun. Fast, 'on-the-fly' diagnostics of big areas is usually done using simplified 'proxy' methods, such as the magnetic line ratio (MLR) method. With the fast development of the Stokes invertion technics and available computational powers, these proxy methods are likely to become obsolete within 10-20 years. However, presently, they remain essential. 

The magnetic line ratio is the most commonly used proxy method for fast intrinsic magnetic field diagnostics in magnetometric observations. It is based on a comparison of the field values observed with two spectral lines ($B_\mathrm{obs,1}$ and $B_\mathrm{obs,2}$) having very close formation depths but different magnetic sensisitivities (i.e. different Lande factors, $g_1$ and $g_2$). The magnetographic signal (Stokes V amplitude) is not directly proportional to the magnetic field due to Zeeman saturation effect and, hence, it is normally possible to relate the $B_\mathrm{obs,1}/B_\mathrm{obs,2}$ ratios to the intrinsic field values 
\cite[e.g.][]{sten73,sane88,sola93,race05,khco07,smso17,beor19}.
The MLR is a very robust and efficient method. However, it has one obvious drawback: it requires observations of at least two lines. Furthermore, the reliability of this method is affected if the chosen lines have slightly different formation depths, or slightly different sensitivities to the temperature or density perturbations. 

\citet{gore18} have suggested a new method for the intrinsic magnetic field (or the filling factor) requiring observations of only one line. In that method $B_\mathrm{real}$ values are evaluated using the widths of Stokes V peaks (Stokes V width method, SVW hereafter). They compared this new method with the classical MLR method for the magneto-convective model of the photosphere and found that, at least for the widely used Fe~I 6301.5$\ang$ and 6302.5$\ang$ lines (6301/6301 pair, hereafter), this method is more reliable than MLR. Thus, on average, SVW had errors smaller than MLR. However, this could be expected, taking into account that 6301.5 and 6302.5 lines have rather different formation depths \citep{khco07}. What is more important is that SVW did not show saturation at higher field values (1--2~kG), typical for MLR.  

In this paper, we explore this new method by extending it to new lines often used for solar magnetic field measurements: optical Fe~I 5247.1$\ang$ and 5250.2$\ang$ lines (5247/5250 pair, hereafter), and near-infrared Fe~I 15648.5$\ang$ and 15652.9$\ang$ pair (15648/15652 pair, hereafter). In Section~2, we calculate calibration curves for these lines, linking the intrinsic magnetic field values with the widths of blue peak of Stokes V components for different Stokes I widths. In Section~3, we test the reliability of this method using the synthetic profiles of the considered lines obtained from MHD magneto-convective simulations of the photosphere.

\section{Synthetic Stokes profiles and calibration curves}

In order to understand the effect of the magnetic field and thermodynamic conditions in the photosphere on the Stokes profiles, we calculate profiles of the six chosen spectral lines for different vertical magnetic field strengths $B_z$ and isotropic micro-turbulent velocities $V_{mic}$. We consider a plane-parallel model of the photosphere with the normal direction ($Z$) parallel to the line-of-sight (LOS). The magnetic field strength, determining the Zeeman effect, is varied in the range 0--2~kG, while the micro-turbulent velocities, affecting the line width, were varied between 0 and 1$\times$10$^5$~m/s, which is the Alfven speed for the magnetic field of 1~kG at the photospheric level. These values are set constant with depth. All other parameters (density, temperature, chemical composition and ionisation) vary with depth according to the VAL C model \citep{vere73}.

\begin{figure}[ht!]    
\centerline{\includegraphics[width=0.40\textwidth,clip=]{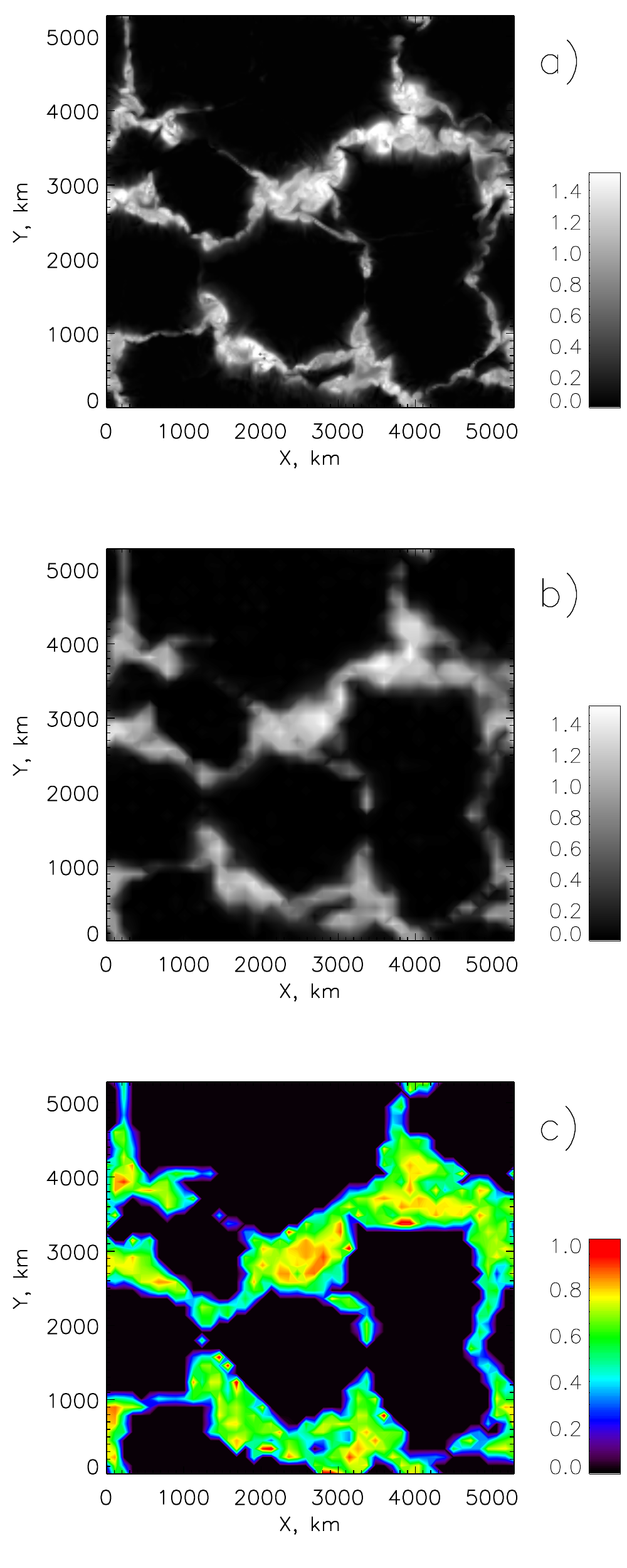}}
\caption{Magnetic field taken from MHD simulations: (a) line-of-sight magnetic field (in kG) with original resolution (5.5~km, please note, that the plots does not reflect the full resolution), (b) line-of-sight magnetic field (in kG) measured using Zeeman splitting in Fe~I~6301$\ang$ line with reduced spatial resolution (110~km), (c) magnetic filling factor corresponding to the spatial resolution of 110~km. The magnetic filling factor is shown only for $|B_\mathrm{obs}|$ field values greater than 100~G.}
\end{figure}

The Stokes profiles are calculated using the NICOLE code \citep{sone00,sone11}. The atomic parameters for the chosen lines are shown in Table~1\footnote{http://www.nist.gov/pml/data/asd.cfm}. The profiles are calculated for the 64$\times$64 parametric grid of $[B,V_{mic}]$. The obtained profiles are used to measure the required parameters: Stokes I width $W_I$ (its full width at half maximum), Stokes V amplitude $V$ (amplitude of Stokes V blue wing normalised on the Stokes I intensity of the continuum), Stokes V width $W_V$ (the full width of Stokes V blue wing at its half maximum), and the splitting between centres of mass of the $I+V$ and $I-V$ profiles, $\Delta \lambda_H$ (Zeeman splitting, hereafter) (Figure~1). Then, our parametric grid $[B,V_{mic}]$ is transformed to $[B,W_I]$, making it possible to link observable parameters -- $W_I$, $V$, $W_V$, $\Delta \lambda_H$ with the magnetic field characteristics.

Increase of magnetic field affects three observables: $V$, $\Delta \lambda _H$, and $W_V$ show noticeable increase. Because the $V$ values for lines with higher Lande factor $g$ saturate faster than in the lines with lower $g$, the magnetic line ratios $R_{MLR} = V_{lower\;g}/V_{higher\;g}$ also increase. Importantly, these parameters are proxies for different magnetic field characteristics. The Stokes V amplitude $V$ and related to it Zeeman splitting $\Delta \lambda _H$ are defined by the spatially-average magnetic field (or a magnetic flux per observational pixel), while the magnetic line ratio $V_{lower\;g}/V_{higher\;g}$ and Stokes V width $W_V$ are proxies of the intrinsic magnetic field strength \cite[see Figures 2-4 in][]{gore18}.

Let us consider the behaviour of the calibration curves, $R_{MLR}(B)$ and $W_V(B)$, for different line widths $W_I$. These parameters are plotted in Figure~2. 

The calibration curves for MLR are very similar to those calculated in previous studies \cite[e.g.][]{khco07,smso17}. It can be seen that both characteristics, MLR and SVW, are proxies of $B$. However, the non-saturation ranges, where they can be effectively used for $B$ estimations, and their sensitivities to the turbulent broadening are very different. As can be expected, in each pair, the line with lower Lande factor saturates at higher $B$ values. For this reason, we will compare SVW calibration curves of the lower $g$ lines with the MLR calibration curves in each pair. 

The MLR curves for the 5247/5250 pair (Figure 2c) can be used in the range from about 100--200~G to 1100--1800~G, depending on the Stokes I profile width. The wider the line, the higher is the $B$ value when MLR saturates, because the condition for $V(B)$ saturation is $\Delta \lambda_H > W_I$. SVW for the 5247 line can be used from 200--300~G (Figure 2 a-b), however, it saturates at larger values of $B$, 1700--2200~G, depending on the Stokes I width.

The 6301/6302 pair appears to have widest range of applicability, compared to the two other pairs. The SVW calibration curves for this pair show no saturation within the considered range, while the MLR calibration curves show very mild saturation around 2~kG. Furthermore, both MLR and SVW calibration curves for this pair show very weak dependence on the widths $W_I$, which, in principle, should result in smaller errors, because the error of $W_I$ measurements is unimportant.

The MLR for the 15648/15652 pair (Figure 2i) has a smaller range of applicability, it saturates at lower $B$ values, simply because $W_I$, which is determined by the Doppler effect, is proportional to $\lambda$, while Zeeman splitting is $\Delta \lambda_H \sim\lambda^2$. Thus, MLR for this pair can be used between 0 and 700--800~G, depending on the profile width $W_I$. Again, the SVW range for the 15652 line is bigger (Figure 2 g-h), it can be used from $\sim$~200~G to 800--1300~G, depending on $W_I$.

Therefore, based purely on the analysis of MLR and SVW calibration curves the 6301/6302 pair appears to be best both for MLR and SVW analysis, compared to the two other pairs, 5247/5250 and 15648/15652. However, these calibration curves are calculated using simplified photospheric models with vertically-uniform magnetic field $B$ and zero macroscopic LOS velocities $V_{LOS}$. Presence of vertical gradients of $B$ (which can be very substantial in the photosphere) and Doppler shifts due to $V_{LOS}$ can substantially affect the Stokes profiles and, hence, results of the MLR and SVW analysis. For instance, it is well known that the MLR method for the 6301/6302 pair can yield substantial errors in the presence of LOS field gradient due to the substantial difference in the line formation depths. Therefore, it is necessary to test the SVW and MLR methods for the three considered line pairs using a realistic magneto-convective model of the photosphere.

The calibration curves are used to create tabulated functions (arrays) of $B_\mathrm{real\; ij}=f\left(R_{MLR},W_I\right)$ and $B_\mathrm{real\; ij}=f\left(W_V, W_I\right)$, which can be directly applied to the observational data in order to estimate the intrinsic field values. It should be noted that it can be done only to limited ranges of $R_{MLR}$, $W_V$ and $W_I$, where these functions are increasing monotonically in respect of both input parameters. Therefore, this calibration can only be applied to pixels within these ranges. 

\begin{figure*}[ht!]    
\centerline{\includegraphics[width=0.75\textwidth,clip=]{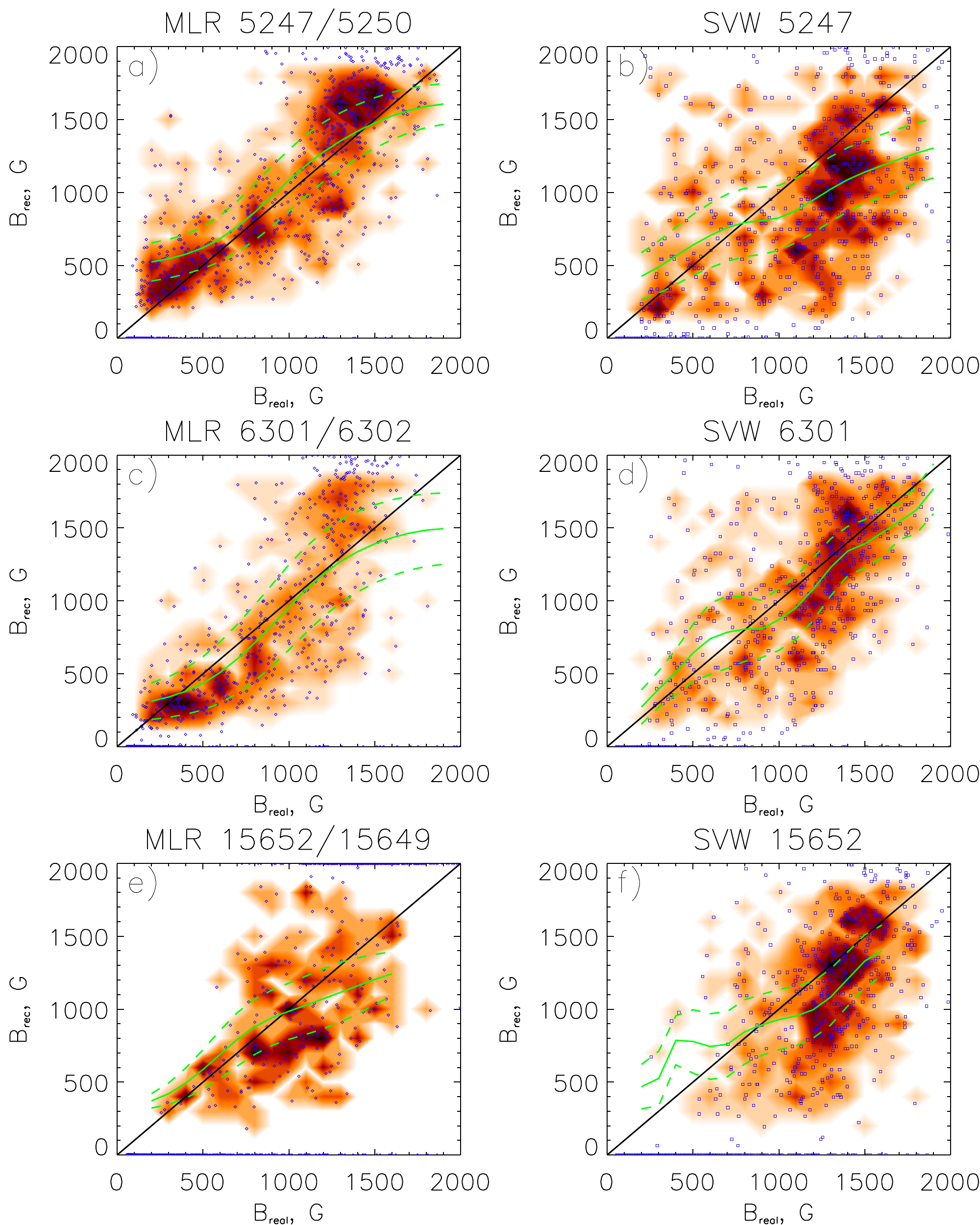}}
\caption{Values of intrinsic field calculated using MLR (left panels) and SVW (right panels) methods compared with the actual field values taken from the model. Panels (a-b), (c-d) and (e-f) correspond to the 5247/5250, 6301/6302 and 15652/15648 pairs, respectively. Solid green lines denote sliding average values (calculated with the 150~G windows); dashed green lines denote $1\sigma$ deviation from the sliding average values.}
\end{figure*}

\section{Testing MLR and SVW methods using an MHD model}

Now let us compare the two methods using one of the high-resolution MHD models of magneto-convection. This three-dimensional models has been developed using the MURAM code \cite[see][]{sche03,shee04,voge05}. Simulations have been performed using a box with a uniform grid of 960$\times$960$\times$400 elements with horizontal step of 5.5~km and vertical step of 5~km. The upper boundary of the domain corresponds to the temperature minimum level. The model consists of two stages: it starts with simulations of convection in the initially plane-parallel photosphere corresponding to the VAL model. Then, a uniform vertical magnetic field is added, and simulations are run for about five convective timescales (corresponding to 50 minutes of physical time). 

The Stokes profiles of selected lines for this model have been calculated using the NICOLE code (see description in Section~2). In order to get magnetograms with limited spatial resolution, we degrade the Stokes I and V cubes corresponding to this MHD model using the same methodology as in \citet{gore18}. The size of 'degraded' pixels is approximately 110--150~km (0.15~arcsec). The magnetic field taken from the MHD model, magnetic field measured using Zeeman effect with reduced spatial resolution, and corresponding magnetic filling factor are shown in Figure~3. 

Obviously, in reality the magnetic field in each degraded pixel has not a single value of intrinsic field but a range of values (although, this range is rather narrow due to magneto-convective collapse, see discussions in \citet{voge05} and \citet{gore18}). Therefore, we define the characteristic filling factor for each pixel using the definition in \citet{gore18}.

Figure~4 compares the actual (i.e. taken from the MHD model) values of the intrinsic magnetic field with values evaluated using the MLR and SVW methods. For SVW estimations we use the line with lower Lande factor $g$ in each pair. For the analysis we only use pixels where the observed field values (i.e. the field values measured using the degraded Stokes cubes) are above 100~G. Obviously, all calibration functions derived in Section~2 have limited applicability ranges (in terms of MLR, SVW and $W_I$  values). Therefore, some 'observed' pixels with values outside of these ranges cannot be analysed. Furthermore, we do not analyse pixels showing abnormal spectral profiles (such as multi-lobe or extremely asymmetric Stokes V profiles). All these failed pixels are assigned default $B_\mathrm{rec}=0$ value. That is why all panels in Figure~4 show numerous pixels at the $B_\mathrm{rec}=0$ axis. The number of failed pixels is apprximately 10$\%$ of the total pixel number. However, their proportion might be bigger if these methods are applied to the photosphere in active regions, with more fragmented velocity field, higher LOS velocity gradients, stronger turbulence and other factors.

The MLR method provides relatively good accuracy for estimations made using the 5247/5250 pair (Figure 4a). In the interval 500-1500~G the typical systematic error is $\sim$100~G. However, below 500~G this method seems to overestimate the intrinsic field, and above 1600--1700~G it shows saturation, as predicted by the calibration (Figure 2). The point spread $\sigma$ is very moderate, varying between 200--300~G in the 200--1800~G interval. In contrast, the SVW estimations using the 5247 line are much less reliable (Figure 4b): the point spread $\sigma$ is around 300--400~G, and there is a substantial systematic error, with the intrinsic field in the range 1000--1500~G being underestimated by 300--500~G. 

The SVW method with the 6301 line appears to be as reliable as MLR with 5247/5250 pair (Figure 4d), the estimations reveal typical point spread of around 200--250~G and the systematic error between 100--200~G. Importantly, these relatively low errors are nearly constant in a wide range of field values, from 200 to 1900~G. The MLR method with the 6301/6302 pair appears to be less reliable (Figure 4c), showing a slightly bigger spread, 250-350~G, and a noticeable systematic error: it seems to quickly saturate above $\sim$1500~G. Still, the MLR estimations with this pair are acceptable in the range below 1200--1300~G. This is consistent with the earlier analysis by \cite{gore18}, although quite surprising, taking into account the difference in line formation depths in this pair \cite[see e.g.][]{khco07}. 

The 15648/15652 pair yields relatively good results with the MLR method in the range below 700-800~G (Figure 4e). The intrinsic field values estimated using both the MLR and SVW methods with this pair of lines have a typical error of around 250~G. Above 1~kG there is a substantial systematic error due to saturation, which is predicted by the calibration (Section~2). The intrinsic field estimations using the SVW method with the 15648 line (Figure 4f) produces results with the point spread of 200--400~G in addition to the systematic error of approximately the same value. More importantly, because this line pair is more sensitive both to the magnetic field variation, and thermal and turbulent line broadening, it produces fewer 'usable' points, particularly below 1~kG. This is because many pixels have observed parameters outside of the acceptable parameter ranges (see Section~2). As the result, the estimations made with this line using the SVW method, should be taken with caution.

\section{Summary and Discussion}

Our study of three pairs of spectral lines commonly used in solar spectropolarimetry shows that all three pairs, in principle, can be used for the intrinsic magnetic field estimations. However, they provide different accuracy and have very different validity ranges.

The 5247/5250 pair can be used both for MLR and SVW measurements. The MLR method is more reliable, yielding estimations with a typical error of 200--300~G, compared with the typical error of 300--350~G for the SVW method. Depending on the width of the Stokes I profile, the MLR method can be used for field values up to 1200--1600~G. The MLR method appears to be less reliable when applied to the 'Hinode pair', 6301/6302. The spread of points is bigger and there is a systematic error -- above 1200~G the magnetic field is underestimated by 300--500~G. At the same time, the SVW method works well: it provides the intrinsic field estimations with the accuracy of about 150--250~G in the range from 200~G to at least 1800~G. The near-infrared pair 15648/15652 provides relatively accurate results, but only for low fields, below 700--800~G. This is not surprising, since the calibration curves show that this pair quickly saturates above $\sim$700~G. 

Overall, our analysis shows that the MLR method with the 5247/5250 pair and the SVW method with the 6301 line provides the most reliable estimations because they do not produce large systematic errors. It should be noted, however, that the analysis is done for one, randomly selected MHD model. Both methods need to be tested further using a number of MHD models of photospheric magneto-convection with different initial magnetic field and different atmospheric structure.

Of course, the reliability of the considered methods depends on the calibration. The calibration used in this and other similar studies \cite[see][]{khco07,smso17,gore18} is based on the vertically-uniform magnetic field. Taking into account that the magnetic field in the photosphere is expected to have substantial vertical gradients, this calibration may result in significant systematic error. This can be overcome by constructing the calibration curves using the MHD models of magneto-convection in the photosphere, similar to those used for testing in this and previous studies.

The main conclusion is that, generally, the SVW method is as good as MLR. The main drawback of the SVW method is that it requires resolved Stokes profiles with resolution of 20--30~m$\ang$ in the green area and 40--50~m$\ang$ in the near-infrared range. At the same time, the SVW method has an important advantage over MLR: it requires only one spectral line and, hence, is not affected by the line formation difference. The latter is practically significant because a number of widely-used instruments observe only one spectral line, or observe a pair not suitable for using the MLR method (such as Hinode/SOT).

Using the SVW method for fast intrinsic magnetic field diagnostics requires spectropolarimentric observations with large field of view and reasonably good spectral resolution, but only moderate spatial resolution.  Forthcoming DKIST \citep{trie16} and future EST \citep{jure19} solar telescopes are expected to provide this type of data in grating spectroscopy mode. The Stokes V Width is a new method and we plan additional tests of it involving high-resolution MHD models in order to investigate the influence of non-LTE effects on the reliability of this method, and by using actual spectropolarimetric data to further compare SVW and MLR. 

\begin{acknowledgements}
MG and PKB are funded by Science and Technology Facilities Council (UK), grant ST/P000428/1. VGL is funded by Taras Shevchenko National University of Kyiv, project 19BF023-03. Simulations have been performed using DiRAC Data Centric system at Durham University, operated by the Institute for Computational Cosmology on behalf of the STFC DiRAC HPC Facility.
\end{acknowledgements}

\end{document}